\documentclass[showpacs, amsmath, amsfonts, amssymb, aps, superscriptaddress]{revtex4}

\usepackage{tabularx}
\usepackage[dvipsnames]{xcolor}
\usepackage{comment}
\usepackage{enumitem}
\usepackage{graphicx}
\usepackage{float}
\usepackage{dcolumn}
\usepackage{bm}
\usepackage{mathrsfs}
\usepackage{amsmath}

\usepackage{amsfonts}
\usepackage{dsfont}

\begin{document}

\title{Instability Analysis of Massive Static Phantom Wormholes via the Spectral Method}
\author{Davide Batic}
\email{davide.batic@ku.ac.ae}
\affiliation{
Mathematics Department, Khalifa University of Science and Technology, PO Box 127788, Abu Dhabi, United Arab Emirates}
\author{Denys Dutykh}
\email{denys.dutykh@ku.ac.ae}
\affiliation{
Mathematics Department, Khalifa University of Science and Technology, PO Box 127788, Abu Dhabi, United Arab Emirates}
\affiliation{Causal Dynamics Pty Ltd, Perth, Australia}

\date{\today}

\begin{abstract}
Using the spectral method, we investigate the scalar and axial quasinormal modes (QNMs) of massive static phantom wormholes. Our results reveal the existence of purely imaginary QNMs that were not identified in previous studies, suggesting potential (in)stabilities as the ratio of the Schwarzschild radius to the wormhole throat varies within a specific range. For scalar perturbations, instabilities arise when this ratio exceeds $1.0$, with the threshold value of $1.0$ itself included. In the case of axial perturbations, the onset of instability occurs at smaller ratios, reflecting the impact of gravitational waves on the wormhole's stability. The findings suggest that the wormhole remains stable when the throat size significantly exceeds the Schwarzschild radius. Our results align with existing literature but offer new insights into the stability conditions of phantom wormholes.
\end{abstract}

\maketitle

\section{Introduction}

Wormholes, hypothetical tunnels connecting distant regions of spacetime, have long fascinated physicists and mathematicians. Flamm discovered the earliest wormhole solution in Einstein's gravity in 1916 \cite{Flamm1916PZ}, though it was not known as a wormhole at that time. In 1928, Weyl proposed a wormhole hypothesis related to electromagnetic field energy, using the term ‘one-dimensional tube’ \cite{Scholz2001}. Einstein and Rosen later rediscovered Flamm's solution, now known as the Einstein-Rosen bridge \cite{Einstein1935PR}. In 1957, Wheeler coined the term ‘wormhole’. The first construction of traversable wormholes was independently done by Ellis \cite{Ellis1973JMP, Ellis1974JMP} and Bronnikov \cite{Bronnikov1973APP}. These were exotic wormholes, requiring a hypothetical ad hoc matter to satisfy certain energy conditions. In 1988, Morris and Thorne \cite{Morris1988AJP} derived a static traversable wormhole, which is a special case of the Ellis-Bronnikov solution. More recently, \cite{Wall2013CQG} used the generalized second law of causal horizons to rule out traversable wormholes connecting two disconnected regions. In 2017, \cite{Gao2017JHEP} constructed short-lived non-exotic traversable wormholes within the AdS/CFT framework, while \cite{Fu2019CQG} provided a perturbative wormhole construction outside this duality.

Among various wormhole solutions, those sustained by phantom energy (a form of exotic matter with negative energy density) have received significant attention due to their unique properties and potential implications for traversable wormholes. In particular, static massive phantom wormholes \cite{Bronnikov1973APP, Ellis1973JMP, Ellis1974JMP, Ellis1979GRG} present an intriguing subclass characterized by their (in)stability properties and interaction with matter and gravitational perturbations. The spacetime geometry of a static massive phantom wormhole can be described by a metric that includes a redshift function dependent on the mass parameter \( M \) and the throat size \( r_0 \). When \( M = 0 \), the metric reduces to the well-known Morris-Thorne wormhole, symmetric with respect to the throat. For non-zero mass, the wormhole exhibits different symmetry properties and gravitational behaviours.

Recent literature has explored various aspects of these wormholes. \cite{BlazquezSalcedo2018PRD} and \cite{Azad2023, Azad2020EPJC, Azad2023PRD} have studied the fundamental properties and stability of phantom wormholes, primarily through numerical simulations using the direct integration method. Specifically, they used the Colsys package \cite{Ascher1979MC}, a collocation method for ordinary differential equations with error estimation and adaptive mesh selection. Their approach involves dividing the space into two regions and matching solutions at a specific point. In contrast, our spectral method captures the global asymptotic behaviour of the solution without dividing the space, resulting in a more comprehensive and efficient analysis of the QNMs, ensuring greater accuracy and robustness in determining the (in)stability of the wormhole geometry.

These studies have revealed that static phantom wormholes can exhibit QNMs, characteristic oscillations that describe how perturbations decay over time. The presence and behaviour of these QNMs are crucial for understanding wormholes' (in)stability under different perturbative conditions. The critical difference between our findings and those of \cite{Gonzalez2009CQG} lies in the conditions under which instability arises. \cite{Gonzalez2009CQG} found universal instability when both the metric and the phantom field are perturbed, while our analysis shows that instability can emerge even when only the phantom field is perturbed, but this instability is not universal and depends on the specific parameters of the wormhole. Furthermore, \cite{BlazquezSalcedo2018PRD} employed a direct integration method to analyze scalar, axial and radial perturbations in phantom wormholes. Their numerical method did not detect any scalar and axial perturbations instability but indicated a radial instability, consistent with findings from \cite{Shinkai2002PRD, Gonzalez2009CQG, Gonzalez2009CQGa}.

Applying the spectral method, which has demonstrated high reliability for determining the QNMs of various black holes and wormholes (see \cite{Batic2024CGG, Batic2024EPJC, Batic2024PRD}), we achieve two main objectives: confirming previous findings and uncovering new insights. More precisely, our analysis of scalar perturbations reveals the emergence of purely imaginary positive QNMs, signalling instabilities when the Schwarzschild radius and the wormhole throat are approximately equal. Notably, the same threshold also marks the onset of instabilities for gravitational perturbations, highlighting a consistent critical behaviour across different types of perturbations.

This paper is structured as follows: Section II presents the theoretical framework and equations governing the scalar and axial perturbations of static massive phantom wormholes. Section III describes the spectral method used for our analysis. Section IV discusses the numerical results, highlighting stability regions and the onset of instabilities. Finally, Section V concludes with a summary of our findings and potential directions for future research.

\section{THE STATIC MASSIVE PHANTOM WORMHOLE: METRIC AND EQUATIONS OF MOTION}

In natural units, where $c = G_N = 1$, the line element describes the spacetime metric of a static massive phantom wormhole \cite{BlazquezSalcedo2018PRD, Azad2023}
\begin{equation}\label{phantomWH}
  ds^2=-e^{f(r)}dt^2+e^{-f(r)}\left[dr^2+(r^2+r_0^2)(d\vartheta^2+\sin^2{\vartheta}d\varphi^2)\right], \quad-\infty<r<+\infty, \quad 0\leq\vartheta\leq\pi, \quad 0\leq\varphi<2\pi,
\end{equation}
where the redshift function is defined as
\begin{equation}\label{f}
f(r)=\frac{C}{r_0}\left[\arctan{\left(\frac{r}{r_0}\right)}-\frac{\pi}{2}\right].    
\end{equation}
Here, the constant $C$ quantifies the degree of symmetry of the wormhole whose mass $M$ is related to $C$ according to $C = 2M$. Hence, for $M = 0$, the wormhole is symmetric with respect to $r = 0$. Moreover, $r_0$ indicates the size of the wormhole's throat. We should also notice that the line element above reduces to the Morris-Thorne wormhole metric in the limit of $M \to 0$. The following sections examine the QNMs modes linked with scalar, axial, and radial perturbations. This analysis is performed by employing the spectral method to solve the corresponding equations derived by \cite{BlazquezSalcedo2018PRD, Azad2023}.

\subsubsection{Scalar perturbations}

In the case of scalar perturbations where the phantom field is perturbed with no backreaction of the metric, \cite{BlazquezSalcedo2018PRD, Azad2023} obtains the following equation
\begin{equation}\label{ODE0}
  e^{f(r)}\frac{d}{dr}\left(e^{f(r)}\frac{dZ_S}{dr}\right)+\left[\omega^2-V_{S}(r)\right]Z_S(r) = 0
\end{equation}
with $f(r)$ given by (\ref{f}) and potential 
\begin{equation}
  V_{S}(r) = e^{2f(r)}\left[\frac{\ell(\ell+1)+1}{r^2+r_0^2}-\frac{(2r-C)^2}{4(r^2+r_0^2)^2}\right].
\end{equation}
By means of the rescalings $x = r/r_0$ and $\mathfrak{c} = 2M/r_0 = r_s/r_0$ representing the ratio between the Schwarzschild radius of a gravitational object with mass $M$ and the size of the wormhole throat, (\ref{ODE0}) becomes
\begin{equation}\label{ODE1}
  \frac{d^2 Z_S}{dx^2}+p_S(x)\frac{dZ_S}{dx}+q_S(x)Z_S(x)=0   
\end{equation}
with 
\begin{equation}\label{pq}
  p_S(x)=\frac{df}{dx}, \quad q_S(x) = \Omega^2e^{-2f(x)}-\frac{\ell(\ell+1)+1}{1+x^2}+\frac{(2x-\mathfrak{c})^2}{4(1+x^2)^2}, \quad f(x)=\mathfrak{c}\left(\arctan{x}-\frac{\pi}{2}\right), \quad \Omega = \omega r_0.
\end{equation}
As a consistency check, notice that as $\mathfrak{c} \to 0$, the above equation goes over into the corresponding equation governing the scalar perturbations of a Morris-Thorne wormhole \cite{Batic2024PRD}. The asymptotic behaviour of the solutions to equation (\ref{ODE1}) can be efficiently deduced using the method outlined in \cite{Olver1994MAA}. For this purpose, we start by observing that in the limit of $x \to +\infty$,
\begin{equation}\label{pqS}
    p_S(x) = \sum_{\kappa=0}^\infty\frac{\mathfrak{f}_\kappa}{x^k} = \mathcal{O}\left(\frac{1}{x^2}\right), \qquad
    q_S(x) = \sum_{\kappa=0}^\infty\frac{\mathfrak{g}_\kappa}{x^k}=\Omega^2+\frac{2\mathfrak{c}\Omega^2}{x}+\mathcal{O}\left(\frac{1}{x^2}\right).
\end{equation}
Given that at least one of the coefficients $\mathfrak{f}_0$, $\mathfrak{g}_0$, $\mathfrak{g}_1$ is nonzero, a formal solution to (\ref{ODE1}) is represented by \cite{Olver1994MAA}
\begin{equation}\label{olvers}
    Z_S^{(j)}(x) = x^{\mu_j}e^{\lambda_j x}\sum_{\kappa=0}^\infty\frac{a_{\kappa,j}}{x^\kappa}, \qquad j \in \{1,2\},
\end{equation}
where $\lambda_1$, $\lambda_2$, $\mu_1$ and $\mu_2$ are the roots of the characteristic equations
\begin{equation}\label{chareqns}
   \lambda^2+\mathfrak{f}_0\lambda+\mathfrak{g}_0=0,\quad
   \mu_j=-\frac{\mathfrak{f}_1\lambda_j+\mathfrak{g}_1}{\mathfrak{f}_0+2\lambda_j}.
\end{equation}
A straightforward computation shows that $\lambda_\pm = \pm i\Omega$. Since the scalar field should behave as $e^{i\Omega x}$ as $x\to +\infty$, we pick the root $\lambda_+$. It results that $\mu=i\mathfrak{c}\Omega$. Hence, the QNM boundary condition at positive space-like infinity can be formulated as
\begin{equation}\label{QNMBCposinf}
    Z_S\underset{{x\to +\infty}}{\longrightarrow} x^{i\mathfrak{c}\Omega}e^{i\Omega x}.
\end{equation}
On the other hand, when $x\to-\infty$, we find that as expected that the asymptotic behavior of $p_S(x)$ is again captured by the first equation in (\ref{pqS}) while the function $q(x)$ behaves at negative space-like infinity as follows
\begin{equation}
    q_S(x)= \sum_{\kappa=0}^\infty\frac{\widehat{\mathfrak{g}}_\kappa}{x^k}=e^{2\pi\mathfrak{c}}\Omega^2 +\frac{2\mathfrak{c}e^{2\pi\mathfrak{c}}\Omega^2}{x}+\mathcal{O}\left(\frac{1}{x^2}\right).
\end{equation}
At this point, we can proceed as above and deduce that the QNM boundary condition at negative space-like infinity can be expressed as
\begin{equation}\label{QNMBCneginf}
    Z_S\underset{{x\to -\infty}}{\longrightarrow} x^{-i\mathfrak{c}e^{\pi\mathfrak{c}}\Omega}e^{ie^{\pi\mathfrak{c}}\Omega x}.
\end{equation}
The application of the Spectral Method relies on the following two steps: First, we transform the radial function $Z(x)$ into a new radial function $\Phi(x)$ such that the QNM boundary conditions (\ref{QNMBCposinf}) and (\ref{QNMBCneginf}) are automatically implemented and $\Phi_S(x)$ is regular as $x\to\pm \infty$. To this aim, we consider the transformation
\begin{equation}\label{AnsatzS}
Z_S(x)=\left(1-\frac{2}{\pi}\arctan{x}\right)^{-i\mathfrak{c}\Omega}\left(1+\frac{2}{\pi}\arctan{x}\right)^{i\mathfrak{c}e^{\pi\mathfrak{c}}\Omega}e^{\frac{2i\Omega x}{\pi}e^{\frac{\pi\mathfrak{c}}{2}\left(1-\frac{2}{\pi}\arctan{x}\right)}\arctan{x}}\Phi_S(x).
\end{equation}
It is gratifying to observe that the above ansatz contains as a special case ($\mathfrak{c} = 0$) the corresponding one used in \cite{Batic2024PRD}. Finally, we introduce the coordinate transformation
\begin{equation}\label{trafoy}
    y=\frac{2}{\pi}\arctan{x},
\end{equation}
which maps the whole real line to the interval $(-1, 1)$. As a result, we end up with the following ordinary differential equation for the radial eigenfunctions, namely
\begin{equation}\label{ODESscalar}
    S_{2,S}(y)\ddot{\Phi}_S(y) + S_{1,S}(y)\dot{\Phi}_S (y) + S_{0,S}(y)\Phi_S (y) = 0,
\end{equation}
where
\begin{equation}\label{S210scalar}
  S_{2,S}(y)=\frac{4}{\pi^2}\cos^4{\left(\frac{\pi y}{2}\right)}, \quad
  S_{1,S}(y) =i\Omega E_{1,S}(y)+F_{1,S}(y),\quad
  S_{0,S}(y)=\Omega^2\Sigma_{2,S}(y)+i\Omega\Sigma_{1,S}(y)+\Sigma_{0,S}(y)
\end{equation}
with
\begin{eqnarray}
  E_{1,S}(y)&=&-\frac{2e^{\frac{\pi\mathfrak{c}}{2}(1-y)}}{\pi^2}\cos^2{\left(\frac{\pi y}{2}\right)}\left[(\pi\mathfrak{c}y-2)\sin{(\pi y)-2\pi y}\right]+\frac{8\mathfrak{c}}{\pi^2}\cos^4{\left(\frac{\pi y}{2}\right)}\left[\frac{e^{\pi\mathfrak{c}}}{1+y}+\frac{1}{1-y}\right], \\
  F_{1,S}(y)&=&\frac{2}{\pi^2}\cos^2{\left(\frac{\pi y}{2}\right)}\left[\mathfrak{c}\pi\cos^2{\left(\frac{\pi y}{2}\right)}-\pi\sin{(\pi y)}\right],
\end{eqnarray}
and
\begin{eqnarray}
\Sigma_{2,S}&=&-\frac{4\mathfrak{c}^2}{\pi^2(1-y^2)^2}\cos^4{\left(\frac{\pi y}{2}\right)}\left[(1-y)^2 e^{2\pi\mathfrak{c}}+2(1-y^2)e^{\pi\mathfrak{c}}+(1+y)^2\right]\nonumber\\
&&+\frac{4\mathfrak{c}}{\pi^2(1+y)}\cos^2{\left(\frac{\pi y}{2}\right)}\left[\left(\frac{\pi\mathfrak{c}}{2}y-1\right)\sin{(\pi y)}-\pi y\right]\left[e^{\frac{\pi\mathfrak{c}}{2}(1-y)}-e^{\frac{\pi\mathfrak{c}}{2}(3-y)}\right]\nonumber\\
&&+\frac{e^{\frac{\pi\mathfrak{c}}{2}(1-y)}}{\pi^2}\left[\pi^2(1-y^2)+\pi y(\pi\mathfrak{c}y-2)\sin{(\pi y)}
-\frac{(\pi\mathfrak{c}y-2)^2}{4}\sin^2{\left(\pi y\right)}\right],\\
\Sigma_{1,S}(y)&=&-8\mathfrak{c}e^{\pi\mathfrak{c}}(1-y)^2\cos^3{\left(\frac{\pi y}{2}\right)}\left\{-2\pi(1+y)\sin{\left(\frac{\pi y}{2}\right)}+\left[\pi\mathfrak{c}(1+y)-2\right]\cos{\left(\frac{\pi y}{2}\right)}\right\}\nonumber\\
&&+4\pi(1-y^2)^2\cos^2{\left(\frac{\pi y}{2}\right)}e^{\frac{\pi\mathfrak{c}}{2}(1-y)}\left[
2(\pi\mathfrak{c}y-2)\cos^2{\left(\frac{\pi y}{2}\right)}+\mathfrak{c}\sin{(\pi y)}-\pi\mathfrak{c} y
\right]\nonumber\\
&&+8\mathfrak{c}(1+y)^2\cos^3{\left(\frac{\pi y}{2}\right)}\left\{2\pi(1-y)\sin{\left(\frac{\pi y}{2}\right)}-\left[\pi\mathfrak{c}(1-y)+2\right]\cos{\left(\frac{\pi y}{2}\right)}\right\},\\
\Sigma_{0,S}(y) &=&\pi^2(1-y^2)^2\cos^2{\left(\frac{\pi y}{2}\right)}\left[(4-\mathfrak{c}^2)\cos^2{\left(\frac{\pi y}{2}\right)}+2\mathfrak{c}\sin{(\pi y)}+4\ell(\ell+1)\right].
\end{eqnarray}

\begin{table}
\caption{Classification of the points $y = \pm 1$ for the relevant functions defined in (\ref{S210scalar}). The abbreviation $z$ ord $n$ stands for zero of order $n$.}
\begin{center}
\begin{tabular}{|c|c|c|c|c|c|}
\hline
$y$      & $S_{2,S}(y)$          & $S_{1,S}(y)$   & $S_{0,S}(y)$\\ \hline
$-1$     & z \mbox{ord} 4        & z \mbox{ord} 2 & z \mbox{ord} 2 \\ \hline
$+1$     & z \mbox{ord} 4        & z \mbox{ord} 2 & z \mbox{ord} 2\\ \hline
\end{tabular}
\label{tableZweiscalar}
\end{center}
\end{table}
Table~\ref{tableZweiscalar} shows that all coefficients entering in (\ref{ODESscalar}) share a common zero of order $2$ at $y = \pm1$. Hence, in order to apply the spectral method, we need to multiply the aforementioned equation by $(1-y^2)^2$ and collect the spectral parameter $\Omega$. This leads to the representation
\begin{equation}\label{TSCH}
  \widehat{L}_{0,S}\left[\Phi_S, \dot{\Phi}_S, \ddot{\Phi}_S\right] +  i\widehat{L}_{1,S}\left[\Phi_S, \dot{\Phi}_S, \ddot{\Phi}_S\right] \Omega +  \widehat{L}_{2,S}\left[\Phi_S, \dot{\Phi}_S, \ddot{\Phi}_S\right] \Omega^2 = 0
\end{equation}
with
\begin{eqnarray}
  \widehat{L}_{0,S}\left[\Phi_S, \dot{\Phi}_S, \ddot{\Phi}_S\right]  &=& \widehat{L}_{00,S}(y)\Phi_S + \widehat{L}_{01,S}(y)\dot{\Phi}_S + \widehat{L}_{02,S}(y)\ddot{\Phi}_S,\label{L0none}\\
  \widehat{L}_{1,S}\left[\Phi_S, \dot{\Phi}_S, \ddot{\Phi}_S\right]  &=& \widehat{L}_{10,S}(y)\Phi_S + \widehat{L}_{11,S}(y)\dot{\Phi}_S + \widehat{L}_{12,S}(y)\ddot{\Phi}_S, \label{L1none}\\
  \widehat{L}_{2,S}\left[\Phi_S, \dot{\Phi}_S, \ddot{\Phi}_S\right]  &=& \widehat{L}_{20,S}(y)\Phi_S + \widehat{L}_{21,S}(y)\dot{\Phi}_S + \widehat{L}_{22,S}(y)\ddot{\Phi}_S,\label{L2none}
\end{eqnarray}
where
\begin{eqnarray}
\widehat{L}_{00,S}(y)&=&-\frac{\cos^2{\left(\frac{\pi y}{2}\right)}}{4(1-y^2)^2}\left[(4-\mathfrak{c}^2)\cos^2{\left(\frac{\pi y}{2}\right)}+2\mathfrak{c}\sin{(\pi y)}+4\ell(\ell+1)\right],\\
\widehat{L}_{01,S}(y)&=&\frac{2\cos^2{\left(\frac{\pi y}{2}\right)}}{\pi(1-y^2)^2}\left[\mathfrak{c}\cos^2{\left(\frac{\pi y}{2}\right)}-\sin{(\pi y)}\right],\quad
\widehat{L}_{02,S}(y)=\frac{4\cos^4{\left(\frac{\pi y}{2}\right)}}{\pi^2(1-y^2)^2},\\
\widehat{L}_{10,S}(y)&=&\frac{\cos^2{\left(\frac{\pi y}{2}\right)}}{\pi^2(1-y^2)^4}\left\{
2\mathfrak{c}e^{\pi\mathfrak{c}}(1-y)^2\cos{\left(\frac{\pi y}{2}\right)}
\left[\frac{\pi\mathfrak{c}}{2}(1+y)-1-\pi(1+y)\sin{(\pi y)}+\left(\frac{\pi\mathfrak{c}}{2}(1+y)-1\right)\cos{(\pi y)}
\right]\right.\nonumber\\
&&\left.-\pi(1-y^2)^2 e^{\frac{\pi\mathfrak{c}}{2}(1-y)}\left[\mathfrak{c}\sin{(\pi y)}+(\pi\mathfrak{c}y-2)\cos{(\pi y)}-2\right]\right.\nonumber\\
&&\left.-2\mathfrak{c}(1+y)^2\left[\pi(1-y)\sin{(\pi y)}-\left(\frac{\pi\mathfrak{c}}{2}(1-y)+1\right)(1+\cos{(\pi y)})\right],
\right\},\\
\widehat{L}_{11,S}(y)&=&\frac{2\cos^2{\left(\frac{\pi y}{2}\right)}}{\pi^2(1-y^2)^3}\left\{
(1-y^2)e^{\frac{\pi\mathfrak{c}}{2}(1-y)}\left[2\pi y-(\pi\mathfrak{c}y-2)\sin{(\pi y)}\right]+4\mathfrak{c}\left[1+y+e^{\pi\mathfrak{c}}(1-y)\right]\cos^2{\left(\frac{\pi y}{2}\right)}\right\},\\
\widehat{L}_{12,S}(y)&=&0,\\
\widehat{L}_{20,S}(y)&=&-\frac{1}{\pi^2(1-y^2)^4}\left\{
4\mathfrak{c}^2\cos^4{\left(\frac{\pi y}{2}\right)}\left[
(1-y) e^{\pi\mathfrak{c}}+1+y\right]^2\right.\nonumber\\
&&\left.-4\mathfrak{c}(1-y^2)\cos^2{\left(\frac{\pi y}{2}\right)}\left[(1+y)e^{\frac{\pi\mathfrak{c}}{2}(1-y)}+(1-y)e^{\frac{\pi\mathfrak{c}}{2}(3-y)}\right]\left[\left(\frac{\pi\mathfrak{c}y}{2}-1\right)\sin{(\pi y)}-\pi y\right]\right.\nonumber\\
&&\left.-(1-y^2)^2 e^{\pi\mathfrak{c}(1-y)}\left[\pi y(\pi\mathfrak{c}y-2)\sin{(\pi y)}-\frac{1}{4}(\pi\mathfrak{c}y-2)^2\sin^2{(\pi y)}+\pi^2(1-y^2)\right]\right\},\\
\widehat{L}_{21,S}(y)&=&0=\widehat{L}_{22,S}(y).
\end{eqnarray}
Table~\ref{tableDreiScalar} shows that all the $\widehat{L}_{ij}$ terms appearing in (\ref{L0none})-(\ref{L2none}) exhibit a regular behavior at the endpoints $y = \pm 1$. Moreover, as a consistency check, we verified that such limits correctly reproduce those displayed in Table~II in \cite{Batic2024PRD} for the case $\mathfrak{c}=0$.

\begin{table}
\caption{Behavior of the coefficients $\widehat{L}_{ij}$ at the endpoints of the interval $-1 \leq y \leq 1$ for the scalar perturbation of the massive phantom wormhole. Here, $\mathfrak{c}=2M/r_0$.}
\begin{center}
\begin{tabular}{ | c | c | c | c | c | c | c | c }
\hline
$(i,j)$  & $\displaystyle{\lim_{y\to -1^+}}\widehat{L}_{ij,S}$  & $\displaystyle{\lim_{y\to 1^-}}\widehat{L}_{ij,S}$  \\ \hline
$(0,0)$ &  $-\frac{\pi^2}{16}\ell(\ell+1)$                             & $-\frac{\pi^2}{16}\ell(\ell+1)$\\ \hline
$(0,1)$ &  $0$                              & $0$\\ \hline
$(0,2)$ &  $0$                              & $0$\\ \hline 
$(1,0)$ &  $0$                              & $0$\\ \hline 
$(1,1)$ &  $-\frac{\pi}{4}e^{\pi\mathfrak{c}}$                             & $\frac{\pi}{4}$\\ \hline 
$(1,2)$ &  $0$                                          & $0$\\ \hline 
$(2,0)$ &  $\frac{\pi\mathfrak{c}}{8}\left[(\pi\mathfrak{c}+2)e^{\pi\mathfrak{c}}+1\right]$            & $\frac{\pi\mathfrak{c}}{8}\left(\pi\mathfrak{c}-2-e^{\pi\mathfrak{c}}\right)$\\ \hline
$(2,1)$ &  $0$                                           & $0$\\ \hline
$(2,2)$ &  $0$                                           & $0$\\ \hline
\end{tabular}
\label{tableDreiScalar}
\end{center}
\end{table} 

\subsubsection{Axial perturbations}

In the case of axial perturbations not coupling to the phantom field, \cite{BlazquezSalcedo2018PRD, Azad2023} obtain the Schr\"{o}dinger-like equation
\begin{equation}\label{ODE2}
  e^{f(r)}\frac{d}{dr}\left(e^{f(r)}\frac{dZ_A}{dr}\right)+\left[\omega^2-V_{A}(r)\right]Z_A(r)=0    
\end{equation}
with $f(r)$ given by (\ref{f}) and potential 
\begin{equation}
  V_{A}(r)=e^{2f(r)}\left[\frac{\ell(\ell+1)-3}{r^2+r_0^2}+\frac{3(2r-C)^2}{4(r^2+r_0^2)^2}\right].   
\end{equation}
By means of the rescalings $x=r/r_0$ and $\mathfrak{c}=2M/r_0=r_s/r_0$, (\ref{ODE2}) becomes
\begin{equation}\label{ODE1A}
  \frac{d^2 Z_A}{dx^2}+p_A(x)\frac{dZ_A}{dx}+q_A(x)Z_A(x)=0   
\end{equation}
with $p_A(x)=p_S(x)$ (see equation (\ref{pq})) and
\begin{equation}\label{q}
  q_A(x)=\Omega^2e^{-2f(x)}-\frac{\ell(\ell+1)-3}{1+x^2}+\frac{3(2x-\mathfrak{c})^2}{4(1+x^2)^2},\quad\Omega=\omega r_0.
\end{equation}
As a consistency check, notice that as $\mathfrak{c} \to 0$, the above equation goes over into the corresponding equation governing the gravitational perturbations of a Morris-Thorne wormhole \cite{Batic2024PRD}. Since the asymptotic behaviours of the coefficient function $p_A(x)$ and $q_A(x)$ are the same as those studied for the scalar case, we can use {\it{mutatis mutandis}} the same ansatz (\ref{AnsatzS}) which encodes simultaneously the correct QNMs boundary condition. If, in addition, we make the variable transformation (\ref{trafoy}), we end up with
\begin{equation}\label{ODEaxial}
    S_{2,A}(y)\ddot{\Phi}_A(y) + S_{1,A}(y)\dot{\Phi}_A (y) + S_{0,A}(y)\Phi_A (y) = 0,
\end{equation}
where $\Phi_A(y)$ is required to be regular at $y=\pm 1$ and 
\begin{equation}\label{S210axial}
S_{2,A}(y)=S_{2,S}(y),\quad
S_{1,A}(y)=S_{1,S}(y),\quad
S_{0,A}(y)=\Omega^2\Sigma_{2,A}(y)+i\Omega\Sigma_{1,A}(y)+\Sigma_{0,A}(y)
\end{equation}
with $\Sigma_{2,A}(y)=\Sigma_{2,S}(y)$, $\Sigma_{1,A}(y)=\Sigma_{1,S}(y)$, and
\begin{equation}
\Sigma_{0,A}(y)=\pi^2(1-y^2)^2\cos^2{\left(\frac{\pi y}{2}\right)}\left[3(\mathfrak{c}^2-4)\cos^2{\left(\frac{\pi y}{2}\right)}-6\mathfrak{c}\sin{(\pi y)}+4\ell(\ell+1)\right].    
\end{equation}
As in the scalar case, all coefficients appearing in (\ref{ODEaxial}) share a common zero of order 2 at $y = \pm 1$. Hence, in order to apply the spectral method, we need to multiply the aforementioned equation by $(1-y^2)^2$ and collect the spectral parameter $\Omega$. This leads to the representation
\begin{equation}\label{TSCHax}
  \widehat{L}_{0,A}\left[\Phi_A, \dot{\Phi}_A, \ddot{\Phi}_A\right] +  i\widehat{L}_{1,A}\left[\Phi_A, \dot{\Phi}_A, \ddot{\Phi}_A\right] \Omega +  \widehat{L}_{2,A}\left[\Phi_A, \dot{\Phi}_A, \ddot{\Phi}_A\right] \Omega^2 = 0
\end{equation}
with
\begin{eqnarray}
  \widehat{L}_{0,A}\left[\Phi_A, \dot{\Phi}_A, \ddot{\Phi}_A\right]  &=& \widehat{L}_{00,A}(y)\Phi_A + \widehat{L}_{01,A}(y)\dot{\Phi}_A + \widehat{L}_{02,A}(y)\ddot{\Phi}_A,\label{L0noneax}\\
  \widehat{L}_{1,A}\left[\Phi_A, \dot{\Phi}_A, \ddot{\Phi}_A\right]  &=& \widehat{L}_{10,A}(y)\Phi_A + \widehat{L}_{11,A}(y)\dot{\Phi}_A + \widehat{L}_{12,A}(y)\ddot{\Phi}_A, \label{L1noneax}\\
  \widehat{L}_{2,A}\left[\Phi_A, \dot{\Phi}_A, \ddot{\Phi}_A\right]  &=& \widehat{L}_{20,A}(y)\Phi_S + \widehat{L}_{21,A}(y)\dot{\Phi}_A + \widehat{L}_{22,A}(y)\ddot{\Phi}_A,\label{L2noneax}
\end{eqnarray}
where all the $\widehat{L}_{ij, A}(y)$'s coincide with the corresponding counterparts of the scalar case except for
\begin{equation}
    \widehat{L}_{00,A}(y)=-\frac{\cos^2{\left(\frac{\pi y}{2}\right)}}{3(1-y^2)^2}\left[3(\mathfrak{c}^2-4)\cos^2{\left(\frac{\pi y}{2}\right)}-6\mathfrak{c}\sin{(\pi y)}+4\ell(\ell+1)\right].
\end{equation}
Moreover, all the $\widehat{L}_{ij,A}$ terms appearing in (\ref{L0noneax})-(\ref{L2noneax}) exhibit a regular behavior at the endpoints $y = \pm 1$ with limits given as in Table~\ref{tableDreiScalar} .

\section{Numerical method}

In order to solve the differential eigenvalue problems \eqref{TSCH} and \eqref{TSCHax} to determine the QNMs along with the corresponding frequencies $\Omega$, we have to discretize the differential operators $\widehat{L}_{j(\kappa)}[\cdot]$ with $j \in \{1,2,3\}$ and $\kappa \in \{S,A\}$ defined in (\ref{L0none})-(\ref{L2none}) and (\ref{L0noneax})-(\ref{L2noneax}), respectively. Since our problem is posed on the finite interval $[-1, 1]$ without any boundary conditions, more precisely, we only require that the functions $\Phi_S(y)$ and $\Phi_A(y)$ be regular functions at $y = \pm 1$, then, it is natural to choose a Tchebyshev-type spectral method \cite{Trefethen2000, Boyd2000}. Namely, we are going to expand the function $y \mapsto \Phi_{S,A}(y)$ in the form of a truncated Tchebyshev series
\begin{equation}\label{eq:exp}
  \Phi_{S,A}(y)=\sum_{k=0}^{N} a_{k,\kappa} T_k(y),
\end{equation}
where $\kappa=S,A$, $N\ \in\ \mathbb{N}$ is kept as a numerical parameter, $\{a_{k,\kappa}\}_{k=0}^{N}\ \subseteq\ \mathds{R}$, and $\{T_k(y)\}_{k=0}^{N}$ are the Tchebyshev polynomials of the first kind
\begin{equation}
    T_k: [-1, 1]\ \longrightarrow\ [-1, 1]\,, \qquad y\ \longmapsto\ \cos\,\bigl(k\arccos y\bigr)\,.
\end{equation}
After substituting expansion \eqref{eq:exp} into the differential equation \eqref{TSCH} or \eqref{TSCHax}, we obtain an eigenvalue problem with polynomial coefficients. In order to translate it into the realm of numerical linear algebra, we employ the collocation method \cite{Boyd2000}. Specifically, rather than insisting that the polynomial function in \( y \) is identically zero (a condition equivalent to having polynomial solutions for the differential problems as per equations \eqref{TSCH}), and \eqref{TSCHax}, we impose a weaker requirement. This involves ensuring that the polynomial vanishes at \( N+1 \) strategically selected points. The number $N+1$ coincides exactly with the number of unknown coefficients $\{a_{k,\kappa}\}_{k=0}^{N}$. For the collocation points, we implemented the Tchebyshev roots grid \cite{Fox1968}
\begin{equation}
  y_k= -\cos{\left(\frac{(2k+1)\pi}{2(n+1)}\right)},\quad k\in\{0, 1,\ldots,N\}.
\end{equation}
In our numerical codes, we also implemented the second option of the Tchebyshev extrema grid
\begin{equation*}
  y_k=-\cos{\left(\frac{k\pi}{n}\right)},\quad k\in\{0, 1,\ldots,N\}.
\end{equation*}
The users are free to choose their favourite collocation points. Notice that we used the roots grid in our computation, and in any case, the theoretical performance of the two available options is known to be absolutely comparable \cite{Fox1968, Boyd2000}.

Upon implementing the collocation method, we derive a classical matrix-based quadratic eigenvalue problem, as detailed in \cite{Tisseur2001}
\begin{equation}\label{eq:eig}
  (M_{0,\kappa} + iM_{1,\kappa}\Omega + M_{2,\kappa}\Omega^2)\bf{a}_\kappa =\bf{0}.
\end{equation}
In this formulation, the square real matrices $M_{j,\kappa}$, each of size $(N+1)\times(N+1)$ for $j=0,1,2$, represent the spectral discretizations of the operators $\widehat{L}_{j,\kappa}[\cdot]$, respectively. The problem \eqref{eq:eig} is solved numerically with the \textsc{polyeig} function from \textsc{Matlab}. This polynomial eigenvalue problem yields \(2(N+1)\) potential values for the parameter \(\Omega\). To discern the physical values of \(\Omega\) that correspond to the black hole's QNM modes, we first overlap the root plots for various values of \(N\) in equation \eqref{eq:exp}, such as \(N=350, 380, 400\). We then identify the consistent roots whose positions remain stable across these different \(N\) values.

In order to reduce the rounding and other floating point errors, we performed all our computations with multiple precision arithmetic that is built in \textsc{Maple} and which is brought into \textsc{Matlab} by the \textsc{Advanpix} toolbox \cite{mct2015}. All numerical computations reported in this study were performed with $400$ decimal digit accuracy.

\section{Numerical results}

We start by comparing our results for the scalar case with those obtained in Figures 5 and 6 by \cite{BlazquezSalcedo2018PRD}, where they set $r_0 = 1$ and plot the real and imaginary parts of the QNMs as a function of the wormhole mass $M$. Notice that our $\mathfrak{c}$ varying on the interval $[0, 2]$ corresponds to $M$ in \cite{BlazquezSalcedo2018PRD} ranging from zero to $1$.  For \(\mathfrak{c} = 0\), the real parts of the QNMs in Figure~7 of \cite{BlazquezSalcedo2018PRD} appear to agree with our Table~\ref{table:1}, despite their plot not providing exact numerical values. In Figure~8 (see \cite{BlazquezSalcedo2018PRD}), which shows the imaginary parts of the QNMs for different $\ell$ and wormhole mass values, we observe agreement with our results for $\mathfrak{c} < 1$ and  $\ell \in \{0,1, \ldots, 4\}$.  Discrepancies arise at $\mathfrak{c} = 1.0$, corresponding to $M = 0.5$ in \cite{BlazquezSalcedo2018PRD}, where we detect instabilities for the fundamental modes with multipole numbers $\ell \in \{1, 2, 4\}$ that they missed due to precision limitations of their direct integration method. Further instabilities for scalar perturbations are detected in the range $1.1 \leq \mathfrak{c} \leq 2$ for $\ell \in \{0, 1, 2, 3, 4\}$. Thus, Tables~\ref{table:1} and \ref{table:2} suggest that a massive static phantom wormhole is stable if $r_0 \gtrsim r_S$. However, since linear instability under scalar perturbations also occurs at larger \(\mathfrak{c}\) values, the system is also expected to be nonlinearly unstable, aligning with the findings of \cite{Gonzalez2009CQG, BlazquezSalcedo2018PRD}. Similar considerations apply to \cite{Azad2023}, who used the same numerical method and reported similar graphs for the scalar case.

Concerning gravitational perturbations, the real parts of the QNMs in Figure~10 of \cite{BlazquezSalcedo2018PRD} appear to agree with our findings in Table 3 when \(\mathfrak{c} \geq 0\). In Figure 11 (see \cite{BlazquezSalcedo2018PRD}), which shows the imaginary parts of the QNMs for different \(\ell\) and wormhole mass values, we observe agreement with our results for the multipole numbers \(\ell \in \{2, 3, 4\}\) and higher values of \(\mathfrak{c}\). However, for \(\mathfrak{c} = 1\) (\(M = 0.5\) in \cite{BlazquezSalcedo2018PRD}), we found an instability for \(\ell = 4\) that went undetected in their study. Moreover, we also found purely imaginary overtones for \(\mathfrak{c} \geq 1.1\), which are missing in \cite{BlazquezSalcedo2018PRD} because they only provided the fundamental tones. Further instabilities for scalar perturbations are detected in the range \(1.1 \leq \mathfrak{c} \leq 2\) for \(\ell \geq 2\). Tables~\ref{table:3} and \ref{table:4} indicate that a massive static phantom wormhole remains stable under scalar and gravitational perturbations, provided $r_0 \gtrsim r_S$. Interestingly, our findings reveal an instability region when $r_0 < r_S$, where $r_S$ denotes the Schwarzschild radius associated with the wormhole's asymptotic mass $M$. Such an instability consistently appears in both scalar and gravitational perturbations. The effect is particularly severe due to the ubiquitous presence of gravitational waves in the universe, which would render the massive phantom wormhole unstable for all $\mathfrak{c} \geq 1$.

Finally, we draw the reader's attention to Tables \ref{table:1}-\ref{table:4}, where 'N/A' indicates that data is not available. This does not mean that QNMs are absent; rather, it signifies that even with $400$ Tchebyshev polynomials, no QNMs could be detected. It is possible that increasing the number of Tchebyshev polynomials might reveal additional QNMs. However, we did not explore this further, as $400$ Tchebyshev polynomials were already at the limit of our computational capacity.

\begin{table}
\centering
\caption{QNM modes for scalar perturbations (spin $s = 0$) of the massive static phantom wormhole are presented in the table below for different values of $\mathfrak{c}$ in the interval $[0,\,1]$. The corresponding results are obtained through our spectral method, utilising $400$ polynomials with a precision of $400$ digits. In this context, $\Omega = \omega r_0$ represents the dimensionless frequency, and $b_0$ characterizes the throat size of the wormhole. The notation 'N/A' indicates data not available.}
\label{table:1}
\vspace*{1em}
\begin{tabular}{||c|c|c|c|c|c|c|c|c||}
\hline\hline
$\ell$ & $n$ & $\mathfrak{c} = 0$ & $\mathfrak{c} = 0.1$ & $\mathfrak{c} = 1$ & $\mathfrak{c} = 1.1$ & $\mathfrak{c} = 1.2$ & $\mathfrak{c} = 1.3$ & $\mathfrak{c} = 1.5$ \\ [0.5ex]
\hline\hline
$0$ & $0$ & $0.6814-0.6178i$ & $0.5852-0.5302i$ & $0.2170-0.1906i$ & $0.2007-0.1757i$ & $0.0000+5.8237i$ & $0.0000+7.5177i$ & $0.1531-0.1327i$ \\
    & $1$ & $0.4672-2.1765i$ & $0.4013-1.8673i$ & $0.0000-0.6106i$ & $0.0000-0.5602i$ & $0.1865-0.1628i$ & $0.1740-0.1515i$ & \mbox{N/A}       \\
\hline
$1$ & $0$ & $1.5727-0.5297i$ & $1.3507-0.4548i$ & $0.0000+24.931i$ & $0.0000+15.638i$ & $0.4337-0.1431i$ & $0.4051-0.1335i$ & $0.3573-0.1176i$ \\
    & $1$ & $1.2558-1.7025i$ & $1.0798-1.4611i$ & $0.5036-0.1667i$ & $0.4663-0.1541i$ & $0.3694-0.4540i$ & $0.3462-0.4233i$ & $0.0000-0.3918i$ \\
    & $2$ & $0.8368-3.2361i$ & $0.7202-2.7749i$ & $0.4251-0.5295i$ & $0.3955-0.4891i$ & $0.0000-0.9276i$ & $0.0000-0.8660i$ & \mbox{N/A}       \\ 
    & $3$ & $0.6334-4.9344i$ & $0.5438-4.2320i$ & $0.0000-0.6991i$ & $0.2985-0.8923i$ & $0.0000-1.3606i$ & \mbox{N/A}       & \mbox{N/A}       \\
\hline
$2$ & $0$ & $2.5467-0.5127i$ & $2.1872-0.4402i$ & $0.0000+20.534i$ & $0.7577-0.1510i$ & $0.0000+7.3149i$ & $0.6588-0.1311i$ & $0.0000+2.0080i$ \\
    & $1$ & $2.3450-1.5725i$ & $2.0148-1.3502i$ & $0.8180-0.1631i$ & $0.7110-0.4626i$ & $0.7050-0.1404i$ & $0.6198-0.4017i$ & $0.5815-0.1157i$ \\
    & $2$ & $1.9478-2.7604i$ & $1.6756-2.3696i$ & $0.7665-0.4998i$ & $0.6254-0.8048i$ & $0.6625-0.4302i$ & $0.0000-0.7424i$ & \mbox{N/A}       \\
    & $3$ & $1.4533-4.2052i$ & $1.2533-3.6068i$ & $0.6715-0.8701i$ & $0.5234-1.1940i$ & $0.5848-0.7480i$ & \mbox{N/A}       & \mbox{N/A}       \\
\hline
$3$ & $0$ & $3.5343-0.5069i$ & $3.0355-0.4353i$ & $1.1369-0.1621i$ & $0.0000+17.789i$ & $0.9803-0.1397i$ & $0.9161-0.1305i$ & $0.0000+1.1773i$ \\
    & $1$ & $3.3901-1.5372i$ & $2.9122-1.3201i$ & $1.0993-0.4917i$ & $1.0533-0.1501i$ & $0.9492-0.4236i$ & $0.8875-0.3958i$ & \mbox{N/A}       \\
    & $2$ & $3.0999-2.6232i$ & $2.6641-2.2527i$ & $1.0264-0.8380i$ & $1.0192-0.4553i$ & $0.8891-0.7219i$ & $0.0000-0.5474i$ & \mbox{N/A}       \\
    & $3$ & $2.6722-3.8235i$ & $2.2992-3.2829i$ & $0.9251-1.2138i$ & $0.9532-0.7760i$ & $0.0000-1.1069i$ & \mbox{N/A}       & \mbox{N/A}       \\
    & $4$ & \mbox{N/A}       & $1.8651-4.4725i$ & $0.8103-1.6282i$ & $0.0000-1.1739i$ & $0.0000-2.0671i$ & \mbox{N/A}       & \mbox{N/A}       \\
\hline
$4$ & $0$ & $4.5271-0.5043i$ & $3.8882-0.4331i$ & $0.0000+21.798i$ & $1.3503-0.1498i$ & $1.2568-0.1394i$ & $1.1747-0.1303i$ & $0.0000+0.9756i$ \\
    & $1$ & $4.4151-1.5227i$ & $3.7925-1.3077i$ & $1.4573-0.1617i$ & $1.3236-0.4524i$ & $1.2324-0.4210i$ & $1.1522-0.3934i$ & $0.0000-0.2952i$ \\
    & $2$ & $4.1899-2.5722i$ & $3.5998-2.2091i$ & $1.4279-0.4884i$ & $1.2710-0.7642i$ & $1.1844-0.7112i$ & $0.0000-0.8608i$ & \mbox{N/A}       \\
    & $3$ & $3.8511-3.6818i$ & $3.3104-3.1620i$ & $1.3699-0.8251i$ & $1.1949-1.0921i$ & $0.0000-0.8923i$ & \mbox{N/A}       & \mbox{N/A}       \\
    & $4$ & \mbox{N/A}       & $2.9327-4.2003i$ & $1.2857-1.1792i$ & $1.1024-1.4436i$ & $0.0000-1.4267i$ & \mbox{N/A}       & \mbox{N/A}       \\ [1ex]
 \hline\hline
 \end{tabular}
\end{table}

\begin{table}
\centering
\caption{QNM modes for scalar perturbations (spin $s = 0$) of the massive static phantom wormhole are presented in the table below for different values of $\mathfrak{c}$ in the interval  $[1.6,\,2]$. The corresponding results are obtained through our spectral method, utilising $400$ polynomials with a precision of $400$ digits. In this context, $\Omega = \omega r_0$ represents the dimensionless frequency, and $b_0$ characterizes the throat size of the wormhole. The notation 'N/A' indicates data not available.}
\label{table:2}
\vspace*{1em}
\begin{tabular}{||c|c|c|c|c|c|c|c|c||}
\hline\hline
$\ell$ & $n$ & $\mathfrak{c} = 1.6$ & $\mathfrak{c} = 1.7$ & $\mathfrak{c} = 1.8$ & $\mathfrak{c} = 1.9$ & $\mathfrak{c} = 2.0$ \\ [0.5ex]
\hline\hline
$0$ & $0$ & $0.1443-0.1249i$ & $0.1364-0.1179i$ & $0.0000-0.0968i$ & $0.0000-0.0923i$ & \mbox{N/A} \\
    & $1$ & \mbox{N/A}       & \mbox{N/A}       & \mbox{N/A}       & \mbox{N/A}       & \mbox{N/A} \\
\hline
$1$ & $0$ & $0.3372-0.1109i$ & \mbox{N/A}       & \mbox{N/A} & \mbox{N/A} & \mbox{N/A} \\
    & $1$ & \mbox{N/A}       & \mbox{N/A}       & \mbox{N/A} & \mbox{N/A} & \mbox{N/A} \\
    & $2$ & \mbox{N/A}       & \mbox{N/A}       & \mbox{N/A} & \mbox{N/A} & \mbox{N/A} \\ 
    & $3$ & \mbox{N/A}       & \mbox{N/A}       & \mbox{N/A} & \mbox{N/A} & \mbox{N/A} \\
\hline
$2$ & $0$ & $0.0000+1.7952i$ & $0.0000+1.6313i$ & $0.0000+1.4996i$ & $0.0000+1.3905i$ & \mbox{N/A} \\
    & $1$ & $0.0000-0.2268i$ & \mbox{N/A}       & \mbox{N/A} & \mbox{N/A} & \mbox{N/A} \\
    & $2$ & \mbox{N/A}       & \mbox{N/A}       & \mbox{N/A} & \mbox{N/A} & \mbox{N/A} \\
    & $3$ & \mbox{N/A}       & \mbox{N/A}       & \mbox{N/A} & \mbox{N/A} & \mbox{N/A} \\
\hline
$3$ & $0$ & $0.0000+1.0887i$ & $0.0000+1.0136i$ & $0.0000+0.9488i$ & $0.0000+0.8922i$ & $0.0000+0.8424i$ \\
    & $1$ & \mbox{N/A}       & \mbox{N/A}       & \mbox{N/A}       & \mbox{N/A}       & $0.0000-0.0598i$ \\
    & $2$ & \mbox{N/A}       & \mbox{N/A}       & \mbox{N/A}       & \mbox{N/A}       & \mbox{N/A}       \\
    & $3$ & \mbox{N/A}       & \mbox{N/A}       & \mbox{N/A}       & \mbox{N/A}       & \mbox{N/A}       \\
\hline
$4$ & $0$ & $0.0000+0.9085i$ & $0.0000+0.8493i$ & $0.0000+0.7981i$ & $0.0000+0.7529i$ & $0.0000+0.7128i$ \\
    & $1$ & \mbox{N/A}       & \mbox{N/A}       & \mbox{N/A}       & \mbox{N/A}       & \mbox{N/A}       \\
    & $2$ & \mbox{N/A}       & \mbox{N/A}       & \mbox{N/A}       & \mbox{N/A}       & \mbox{N/A}       \\
    & $3$ & \mbox{N/A}       & \mbox{N/A}       & \mbox{N/A}       & \mbox{N/A}       & \mbox{N/A}       \\ [1ex]
 \hline\hline
 \end{tabular}
\end{table}

\begin{table}
\centering
\caption{QNM modes for axial perturbations (spin $s = 2$) of the massive static phantom wormhole are presented in the table below for different values of $\mathfrak{c}$ in the interval $[0,\,1]$. The corresponding results are obtained through our spectral method, utilising $400$ polynomials with a precision of $400$ digits. In this context, $\Omega = \omega r_0$ represents the dimensionless frequency, and $b_0$ characterizes the throat size of the wormhole. The notation 'N/A' indicates data not available.}
\label{table:3}
\vspace*{1em}
\begin{tabular}{||c|c|c|c|c|c|c|c|c||}
\hline\hline
$\ell$ & $n$ & $\mathfrak{c} = 0$ & $\mathfrak{c} = 0.1$ & $\mathfrak{c} = 1$ & $\mathfrak{c} = 1.1$ & $\mathfrak{c} = 1.2$ & $\mathfrak{c} = 1.3$ & $\mathfrak{c} = 1.5$ \\ [0.5ex]
\hline\hline
$2$ & $0$ & $1.7377-0.3051i$ & $1.4926-0.2635i$ & $0.5705-0.1299i$ & $0.5309-0.1231i$ & $0.4963-0.1168i$ & $0.4658-0.1109i$ & $0.4145-0.1002i$ \\
    & $1$ & $1.7203-1.0396i$ & $1.4760-0.8929i$ & $0.5056-0.3498i$ & $0.4618-0.3291i$ & $0.4240-0.3116i$ & $0.3913-0.2969i$ & $0.3382-0.2735i$ \\
    & $2$ & $1.5249-2.0305i$ & $1.3095-1.7427i$ & $0.4729-0.6202i$ & $0.4339-0.5698i$ & $0.3995-0.5256i$ & $0.3688-0.4868i$ & $0.0000-0.4182i$ \\
    & $3$ & $1.1699-3.3340i$ & $1.0061-2.8604i$ & $0.4050-1.0010i$ & $0.0000-0.9828i$ & $0.0000-0.9123i$ & $0.0000-0.8501i$ & \mbox{N/A}       \\
    & $4$ & $0.8819-4.8985i$ & $0.7587-4.2018i$ & $0.0000-1.0628i$ & $0.0000-1.4431i$ & \mbox{N/A}       & \mbox{N/A}       & \mbox{N/A}       \\
\hline
$3$ & $0$ & $2.9524-0.4100i$ & $2.5366-0.3531i$ & $0.9687-0.1472i$ & $0.0000+10.0916i$ & $0.8388-0.1287i$ & $0.7854-0.1209i$ & $0.0000+1.4647i$ \\
    & $1$ & $2.8625-1.2591i$ & $2.4579-1.0829i$ & $0.9215-0.4435i$ & $0.8995-0.1374i$ & $0.7984-0.3888i$ & $0.7479-0.3657i$ & $0.6956-0.1076i$ \\
    & $2$ & $2.6606-2.1909i$ & $2.2840-1.8822i$ & $0.8295-0.7437i$ & $0.8557-0.4146i$ & $0.7162-0.6556i$ & $0.6712-0.6185i$ & $0.0000-0.3642i$ \\
    & $3$ & $2.3313-3.2596i$ & $2.0022-2.7986i$ & $0.7135-1.0499i$ & $0.7684-0.6969i$ & $0.0000-0.8727i$ & $0.0000-0.8291i$ & \mbox{N/A}       \\
    & $4$ & $1.9109-4.5290i$ & $1.6432-3.8866i$ & $0.0000-1.4853i$ & $0.6521-0.9803i$ & $0.5999-0.9221i$ & \mbox{N/A}       & \mbox{N/A}       \\
\hline
$4$ & $0$ & $4.0763-0.4491i$ & $3.5018-0.3863i$ & $0.0000+17.1287i$ & $1.2328-0.1425i$ & $1.1488-0.1330i$ & $1.0748-0.1247i$ & $0.0000+1.0420i$ \\
    & $1$ & $3.9846-1.3592i$ & $3.4226-1.1687i$ & $1.3288 -0.1531i$ & $1.2024-0.4299i$ & $1.1209-0.4016i$ & $1.0490-0.3764i$ & \mbox{N/A}       \\
    & $2$ & $3.7971-2.3060i$ & $3.2610-1.9820i$ & $1.2957 -0.4620i$ & $1.1421-0.7250i$ & $1.0653-0.6778i$ & $0.0000-0.7253i$ & \mbox{N/A}       \\
    & $3$ & $3.5086-3.3190i$ & $3.0133-2.8515i$ & $1.2297 -0.7791i$ & $1.0521-1.0348i$ & $0.9828-0.9678i$ & \mbox{N/A}       & \mbox{N/A}       \\
    & $4$ & $3.1230-4.4367i$ & $2.6831-3.8102i$ & $1.1320 -1.1104i$ & $0.0000-1.3184i$ & $0.0000-1.2628i$ & \mbox{N/A}       & \mbox{N/A}       \\ [1ex]
 \hline\hline
 \end{tabular}
\end{table}

\begin{table}
\centering
\caption{QNM modes for axial perturbations (spin $s = 2$) of the massive static phantom wormhole are presented in the table below for different values of $\mathfrak{c}$ in the interval $[1.6,\,2]$. The corresponding results are obtained through our spectral method, utilising $400$ polynomials with a precision of $400$ digits. In this context, $\Omega = \omega r_0$ represents the dimensionless frequency, and $b_0$ characterizes the throat size of the wormhole. The notation 'N/A' indicates data not available.}
\label{table:4}
\vspace*{1em}
\begin{tabular}{||c|c|c|c|c|c|c|c|c||}
\hline\hline
$\ell$ & $n$ & $\mathfrak{c} = 1.6$ & $\mathfrak{c} = 1.7$ & $\mathfrak{c} = 1.8$ & $\mathfrak{c} = 1.9$ & $\mathfrak{c} = 2.0$ \\ [0.5ex]
\hline\hline
$2$ & $0$ & $0.3927-0.0954i$ & \mbox{N/A}       & \mbox{N/A}       & \mbox{N/A}       & $0.0000-0.0552i$ \\
    & $1$ & \mbox{N/A}       & \mbox{N/A}       & \mbox{N/A}       & \mbox{N/A}       & \mbox{N/A}       \\
\hline
$3$ & $0$ & $0.0000+1.3344i$ & $0.0000+1.2303i$ & $0.0000+1.1430i$ & $0.0000+1.0685i$ & $0.0000+1.0039i$ \\
    & $1$ & \mbox{N/A}       & \mbox{N/A}       & \mbox{N/A}       & \mbox{N/A}       & \mbox{N/A}       \\
\hline
$4$ & $0$ & $0.0000+0.9669i$ & $0.0000+0.9025i$ & $0.0000+0.8466i$ & $0.0000+0.7975i$ & $0.0000+0.7540i$ \\
    & $1$ & \mbox{N/A}       & $0.0000-0.2003i$ & \mbox{N/A}       & \mbox{N/A}       & \mbox{N/A}       \\ [1ex]
 \hline\hline
 \end{tabular}
\end{table}

\section{Conclusions and outlook}

Using the Spectral Method, we have analyzed the scalar and gravitational QNMs of massive static wormholes supported by a massless phantom field. The wormhole metric depends on the throat parameter $r_0$ and the symmetry parameter $C = 2M$. Due to the symmetry $\omega(C) = \omega(-C)$ \cite{BlazquezSalcedo2018PRD}, we focused on wormholes with nonnegative mass. The Spectral Method, known for its efficiency in studying quasinormal frequencies of various black holes and wormholes \cite{Mamani2022EPJC, Batic2024EPJC, Batic2024PRD, Batic2024CGG}, was employed to obtain QNMs.

We calculated the fundamental modes and overtones across various multipole numbers $\ell$ for both scalar and gravitational perturbations. Our results demonstrate that the massive static phantom wormhole is linearly unstable when $r_0 \lesssim r_S$, where $r_S$ represents the Schwarzschild radius corresponding to the wormhole's asymptotic mass $M$. This instability is observed consistently in both the scalar and gravitational cases. Last but not least, such unstable modes went undetected in earlier studies \cite{BlazquezSalcedo2018PRD, Azad2023}. In particular, the instability under axial perturbations is quite striking due to the ubiquitous presence of gravitational waves in the universe. Based on these linear instabilities, we also expect nonlinear instability, consistent with the findings of \cite{Gonzalez2009CQG, BlazquezSalcedo2018PRD}.

Our study opens several avenues for future research. First, a deeper investigation into the nonlinear stability of these wormholes is necessary, particularly considering the significant influence of gravitational perturbations. Extending the analysis to include other types of perturbations, such as polar perturbations, could provide a more comprehensive understanding of the stability landscape. Additionally, exploring wormholes with different types of exotic matter beyond phantom fields may yield new insights into traversable wormhole configurations. Finally, advancing numerical techniques, possibly through high-performance computing, could overcome current limitations and uncover more complex stability characteristics, further connecting theoretical predictions with potential astrophysical observations.

\section*{Code availability}

All analytical computations reported in this manuscript have been rechecked in the computer algebra system \textsc{Maple}. Two \textsc{Maple} sheets corresponding to the extreme and non-extreme cases can be found in supplementary materials \emph{and} in the repository below. The discretization of differential operators \eqref{L0none}-\eqref{L2none} and \eqref{L0noneax}-\eqref{L2noneax} using the Tchebyshev-type spectral method is equally performed in \textsc{Maple} computer algebra system. Finally, the numerical solution of the resulting quadratic eigenvalue problem \eqref{eq:eig} is performed in \textsc{Matlab} software using the \texttt{polyeig} function. All these materials are freely accessible at the following repository

\begin{itemize}
    \item \url{https://github.com/dutykh/EllisBronnikov/}
\end{itemize}

\section*{Acknowledgements}

This publication is based upon work supported by the Khalifa University of Science and Technology under Award No. FSU$-2023-014$.


\end{document}